\documentclass[twocolumn, superscriptaddress, amsmath, amssymb]{revtex4}

\usepackage{graphicx}% Include figure files
\usepackage{dcolumn}% Align table columns on decimal point
\usepackage{bm}% bold math

\begin{document}

\title{Quasi 2D electronic states with high spin-polarization in centrosymmetric MoS$_2$ bulk crystals}

\author{Mathias Gehlmann}
\affiliation{PGI-6, Forschungszentrum J\"ulich GmbH, D-52425 J\"ulich, Germany}
\author{Gustav Bihlmayer}
\affiliation{PGI-1/IAS-1, Forschungszentrum J\"ulich GmbH and JARA, D-52425 J\"ulich, Germany}
\author{Irene Aguilera}
\affiliation{PGI-1/IAS-1, Forschungszentrum J\"ulich GmbH and JARA, D-52425 J\"ulich, Germany}
\author{Ewa Mlynczak}
\affiliation{PGI-6, Forschungszentrum J\"ulich GmbH, D-52425 J\"ulich, Germany}
\affiliation{Faculty of Physics and Applied Computer Science, AGH University of Science and Technology, al. Mickiewicza 30, 30-059 Krak\'{o}w, Poland}
\author{Markus Eschbach}
\affiliation{PGI-6, Forschungszentrum J\"ulich GmbH, D-52425 J\"ulich, Germany}
\author{Sven D\"oring}
\affiliation{PGI-6, Forschungszentrum J\"ulich GmbH, D-52425 J\"ulich, Germany}
\author{Pika Gospodaric}
\affiliation{PGI-6, Forschungszentrum J\"ulich GmbH, D-52425 J\"ulich, Germany}
\author{Stefan Cramm}
\affiliation{PGI-6, Forschungszentrum J\"ulich GmbH, D-52425 J\"ulich, Germany}
\author{Beata Kardynal}
\affiliation{PGI-9, Forschungszentrum J\"ulich GmbH, D-52425 J\"ulich, Germany}
\author{Lukasz Plucinski}
\email[Correspondence and requests should be addressed to: \\ ]{ l.plucinski@fz-juelich.de }
\affiliation{PGI-6, Forschungszentrum J\"ulich GmbH, D-52425 J\"ulich, Germany}

\author{Stefan Bl\"ugel}
\author{Claus M. Schneider}
\affiliation{PGI-6, Forschungszentrum J\"ulich GmbH, D-52425 J\"ulich, Germany}

\begin{abstract}

Time reversal dictates that nonmagnetic, centrosymmetric crystals cannot be spin-polarized as a whole. 
However, it has been recently shown that the electronic structure in these crystals can in fact show regions of high spin-polarization, as long as it is probed locally in real and  in reciprocal space.
In this article we present the first observation of this type of compensated polarization in MoS$_2$ bulk crystals. 
Using spin- and angle-resolved photoemission spectroscopy (ARPES) we directly observed a spin-polarization of more than 65\% for distinct valleys in the electronic band structure. 
By additionally evaluating the probing depth of our method we find that these valence band states at the $\overline{\text{K}}$~point in the Brillouin zone are close to fully polarized for the individual atomic trilayers of MoS$_2$, which is confirmed by our density functional theory calculations. 
Furthermore, we show that this spin-layer locking leads to the observation of highly spin-polarized bands in ARPES since these states are almost completely confined within two dimensions.
Our findings prove that these highly desired properties of MoS$_2$ can be accessed without thinning it down to the monolayer limit.

\end{abstract}

\flushbottom
\maketitle

\thispagestyle{empty}

\section*{Introduction}

Two-dimensional materials are considered excellent candidates for next-generation electronic devices that could overcome the restrictions of classical, Si-based electronics.
Apart from the possibility of a reduced size compared to state of the art transistor devices, thinning down materials like molybdenum disulphide (MoS$_2$) gives rise to new phenomena that might allow entirely different approaches for applications such as spintronics, valleytronics, solar cells, or optical sensors \cite{doi:10.1038/nature12385}.

\begin{figure*}
\includegraphics[width=1\linewidth]{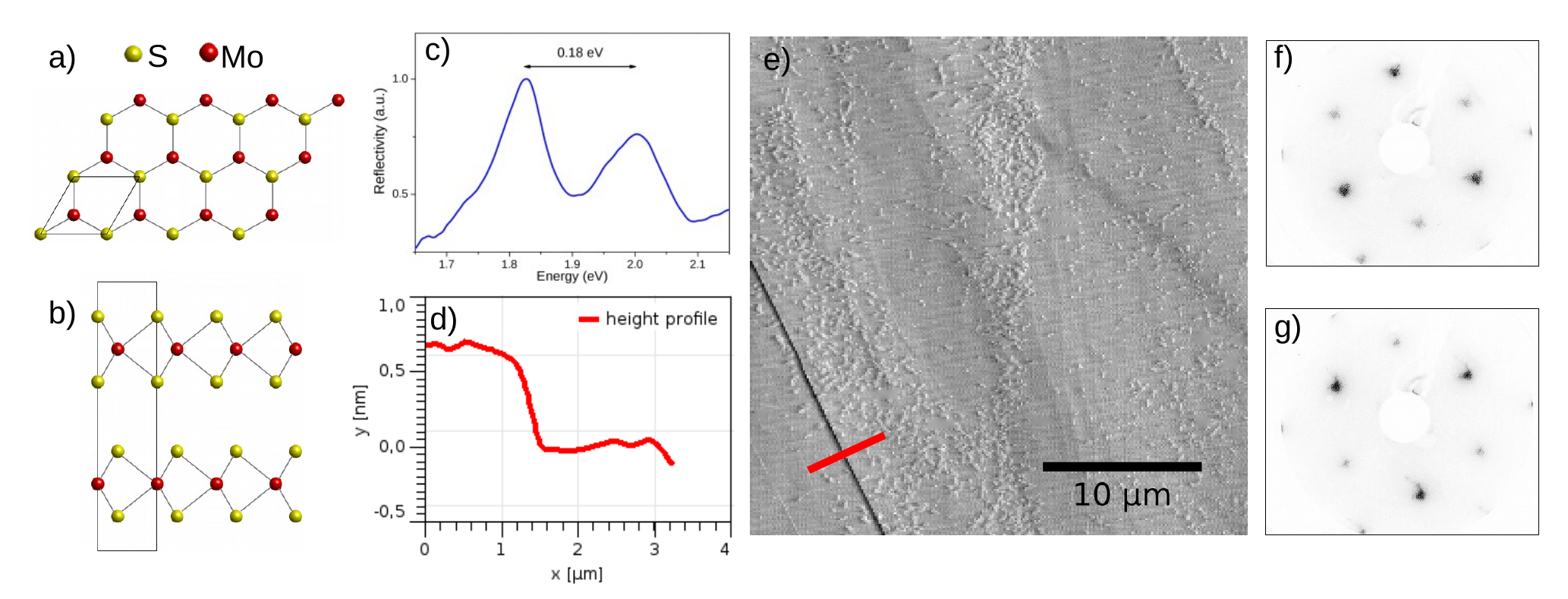}
\caption{\label{fig:sample} 
a) Top view of a MoS$_2$ monolayer, 
b) side view of the 2H crystal structure of bulk MoS$_2$, 
c) optical reflectivity spectrum showing exciton transitions characteristic of 2H-MoS$_2$,  
d) AFM height profile along the red line in AFM image,
e) AFM amplitude image with single monolayer step edge (in the bottom left) separating two large, atomically flat areas. The smaller speckles are adsorbants from ambient air exposure, 
f-g) LEED images taken at two positions on the sample $\approx$~1~mm apart showing a reversing of the threefold symmetry.
}
\end{figure*}

Time reversal dictates that electronic states in nonmagnetic, centrosymmetric crystals cannot be spin-polarized. 
However, a recent publication by Zhang and Liu et al. \cite{nature.phys.10.387} introduces a ``hidden spin-polarization'' resulting from specific site asymmetries instead of the crystal space group.
MoS$_2$ is a model material which exhibits such effects.
It is a layered crystal that consists of two-dimensional S-Mo-S atomic trilayers with broken inversion symmetry within the plane. 
In A-B stacking these trilayers recover inversion symmetry [Fig.~\ref{fig:sample}~b)] and therefore the bulk shows a centrosymmetric space group \cite{PhysRevLett.108.196802,10.1038/nnano.2012.95}.
Spin-polarization in MoS$_2$ is zero when averaging in real or reciprocal space. 
Nevertheless, as it was shown experimentally for other materials, electronic states resolved in space and in momentum can be spin-polarized \cite{PhysRevLett.105.076804,10.1038/nphys3105}.
Angle resolved photoemission spectroscopy (ARPES) meets both of these conditions due to its surface sensitivity and momentum-resolution capabilities and was therefore chosen a tool for the experimental study of the hidden spin-polarization in MoS$_2$. 

The majority of electronic structure studies of MoS$_2$ focuses on quasi freestanding trilayers, which are usually referred to as one monolayer of MoS$_2$, and its integration into novel two-dimensional devices \cite{doi:10.1021/ar500274q,doi:10.1021/nl4042824,PhysRevLett.111.106801}. 
One of the advantageous properties of a freestanding MoS$_2$ monolayer is its direct band gap of about 1.8 eV at the $\overline{\text{K}}$~point, while bulk MoS$_2$ has a smaller, indirect band gap with the valence band maximum (VBM) at the $\Gamma$~point \cite{PhysRevLett.111.106801,PhysRevLett.105.136805}.
Furthermore, at the $\overline{\text{K}}$~point the topmost valence band forms a distinct valley, that makes MoS$_2$ a candidate in the novel field of valleytronics \cite{10.1038/nphys2942}.
In MoS$_2$, as in other transition metal dichalcogenides (TMDCs), the valley is split into two subbands. 
These split states are usually considered to be highly spin-polarized perpendicular to the surface in a freestanding monolayer, but unpolarized in bulk MoS$_2$ \cite{PhysRevLett.108.196802,Suzuki_Sakano,PhysRevLett.111.026601}. 
However, the local nature of spin-orbit coupling (SOC) leads to a locking of the spin to layers.  
An indication for this spin-layer locking is the high circular polarization dependence of the photoluminescence for inversion-symmetric TMDC films, e.g. consisting of an even number of layers or bulk-like crystals \cite{10.1038/nnano.2012.96,10.1038/nphys2942,10.1038/nnano.2012.95}. 
Recently Liu et al. \cite{PhysRevLett.114.087402} derived from first principle calculations that the observed circular polarization is a consequence
of this hidden spin-polarization. 
Consistently with this report, our findings for bulk MoS$_2$ and a recent publication \cite{10.1038/nphys3105} concerning the similar layered crystal WSe$_2$ show that a spin-polarization can be observed, as long as the probing tool focuses on a specific layer in the bulk. 
By the same mechanism, a significant spin-polarization has been measured in the bulk states of Bi \cite{PhysRevLett.105.076804}.

In this article we present a combined spin-resolved ARPES (spin-ARPES) and density functional theory (DFT) study of a cleaved MoS$_2$ single crystal surface.
We focus on the valence band around the $\overline{\text{K}}$~point and in both experiment and theory. 
Specifically,we can show that these electronic states in the bulk material are almost completely confined within the plane of the layers and highly spin-polarized.

\section*{Results and Discussion}
\label{res}

The spin-polarization of the states of MoS$_2$ is hidden by its alternating sign from one layer to the next. 
Although low energy photoemission is sufficiently surface sensitive to probe it, an atomically flat surface (A or B layer terminated) is required for our experiment. 
The amplitude image of an atomic force microscope (AFM) in Fig. \ref{fig:sample}~e) shows the typical surface topography of a cleaved MoS$_2$ single crystal and that the scotch tape method produces atomically flat areas at the MoS$_2$ surface that largely exceed the maximum field of view of the AFM of 30~$\mu\text{m}$.
The height profile cut in Fig. \ref{fig:sample}~d) reveals a step of $\approx$~0.6~nm height, which fits a single monolayer step edge very well. 
The numerous small speckles that are visible within the terraces have a much lower height of $\approx$~0.2~nm. Very likely these are adsorbants from air, since the sample has been exposed to ambient pressure during these measurements. 
However, we have no direct feedback for the microscopic position of the beam spot in the ARPES experiments and have to rely on the quality of the spectra as an indication for a flat area. 
Mahatha et al. \cite{0953-8984-24-47-475504} reported an inhomogeneous band bending in ARPES from areas with high step densities, which we also observed in other areas of our samples. 

Another consideration is concerned with the possible existence of a non-centrosymmetric 3R-MoS$_2$ phase in our samples. 
To rule out the 3R-MoS$_2$ phase as the origin of the spin-polarization that is the central focus of this article, we performed optical reflectivity measurements which are known to show distinguishable exciton transitions for 2H-MoS$_2$ and 3R-MoS$_2$ \cite{Suzuki_Sakano, doi:10.1080/00018736900101307, 0022-3719-5-24-016}.
Figure \ref{fig:sample}~c) shows a reflectivity spectrum from our sample.
The observed separation of the two exciton transitions perfectly matches the expected value of 0.18~eV for 2H-MoS$_2$ compared to 0.14~eV for the 3R-phase. 
This observation is consistent with the reversal of the symmetry in the LEED images in Figure \ref{fig:sample}~f)~and~g) which corresponds to two terraces of the 2H-phase, as they were observed with the AFM, separated in height by an odd number of monolayers. 
From these two evidences, we conclude that in the mineral crystals that were used for our experiments, the centrosymmetric 2H-phase is clearly dominant.

\begin{figure*}
\includegraphics[width=0.9\linewidth]{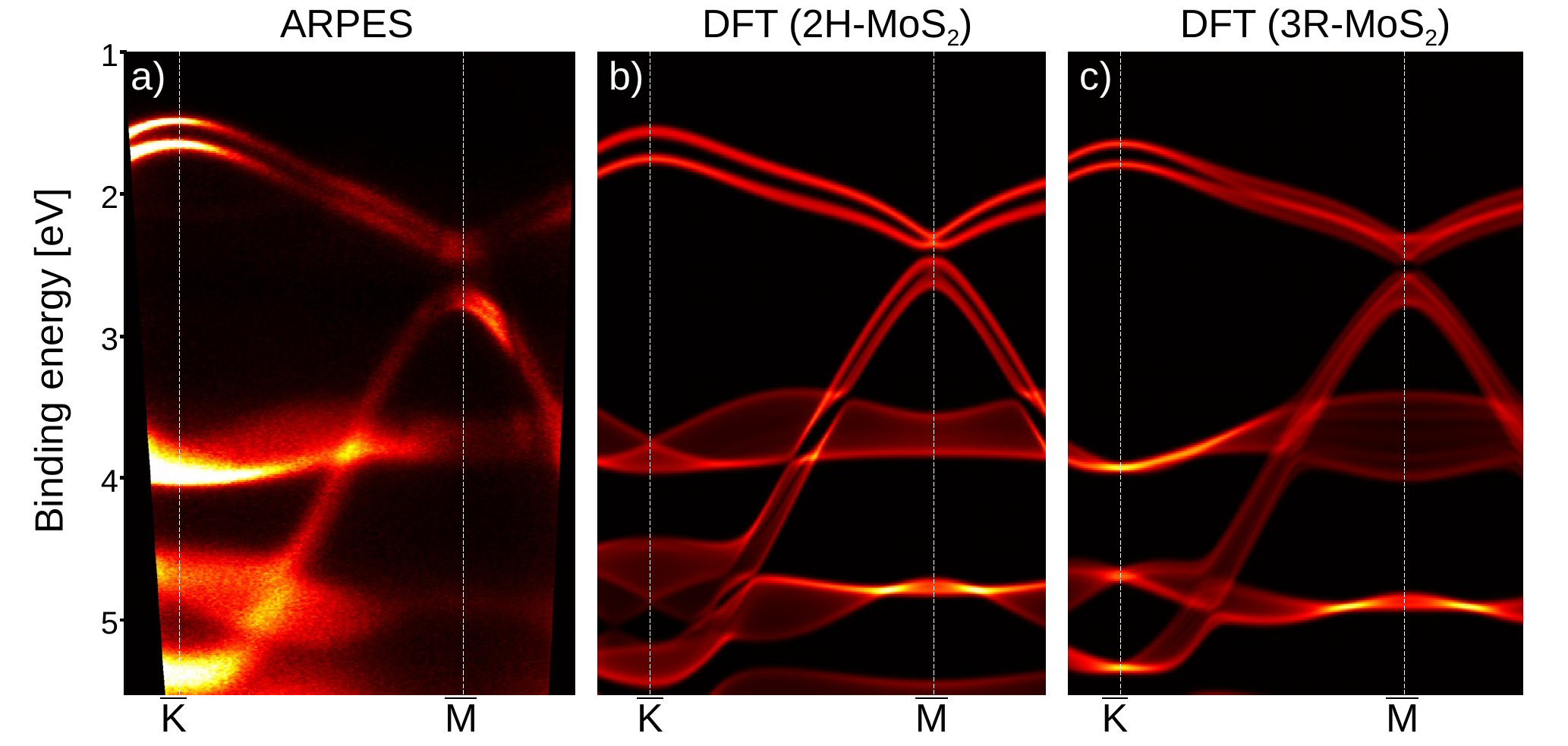}
\caption{\label{fig:all_sbs} 
a) ARPES spectrum along $\overline{\text{KM}}$ direction measured with h$\nu$~=~21.2~eV using He-VUV source and 
 the corresponding calculation of bulk projected band structure for 2H-MoS$_2$ in b) and 3R-MoS$_2$ in c).
 In the calculated maps bright areas indicate sharp bands with little out-of-plane dispersion, darker bands a broader projection.
 In b) and c) the Fermi energy is shifted for comparison with a).
}
\end{figure*}

Figure \ref{fig:all_sbs} depicts a direct comparison of the measured ARPES data with the calculated bulk projected band structure along the $\overline{\text{KM}}$~direction for both 2H-MoS$_2$ and 3R-MoS$_2$.
The same comparison is shown for the $\overline{\Gamma\text{K}}$~direction in Fig. \ref{fig:ARPES_suppl} of the supplementary section.
Although our experiment is extremely surface sensitive we find a very good agreement of the ARPES spectra (in particular at the $\overline{\text{M}}$~point, where the differences are most pronounced) with the calculated bulk band structure of 2H-MoS$_2$, which supports the other evidences that we measured the centrosymmetric crystal phase. 
Finding that the bulk band structure matches ARPES spectra has been reported in several studies \cite{0953-8984-24-47-475504,PhysRevB.86.115105,Suzuki_Sakano,PhysRevB.91.235202} and is not surprising for this class of materials since the layers do not give rise to dangling bonds in the (0001) direction and show almost no surface reconstruction.

\begin{figure*}
\includegraphics[width=1\linewidth]{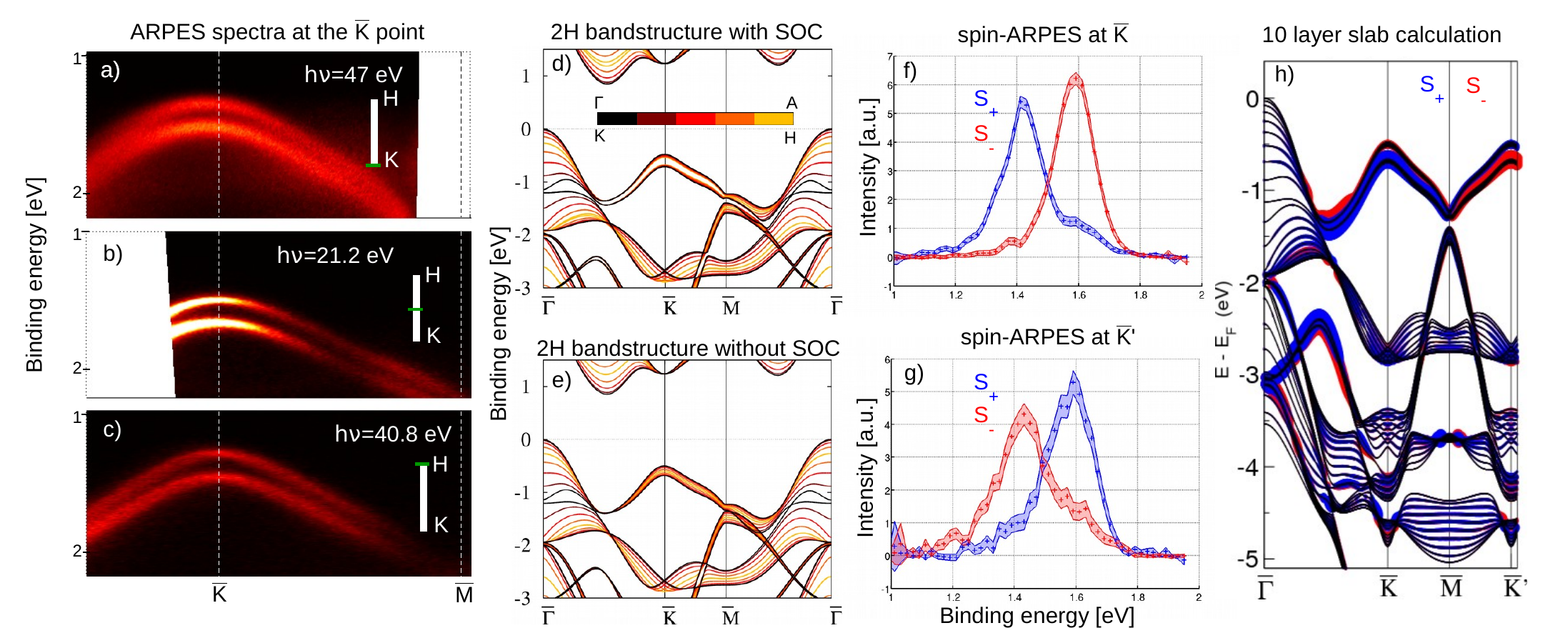}
\caption{\label{fig:splitting_spinpol} 
a-c) ARPES spectra of the VBM at the $\overline{\text{K}}$~point taken at different photon energies to map the k$_\perp$ dependence, 
d-e) DFT calculation of the bulk band structure of MoS$_2$ with and without SOC, color coded for different values of k$_\perp$,
f-g) spin-ARPES spectra of the valence band at the $\overline{\text{K}}$/$\overline{\text{K}}$'~point. These spectra are corrected for the spin-detector efficiency and the non-zero emission angle. The filled area represents the statistical error,
h) 10-layer slab calculation of MoS$_2$ band structure. The size of the red and blue circles represent the out-of-plane spin-polarization in the topmost monolayer.
}
\end{figure*}

At the $\overline{\text{K}}$~point the bulk projection, which is the line KH of the 3D bulk band structure projected into the surface plane, reveals two very sharp subbands of the topmost valence band. 
Although our bulk calculations are in good overall agreement with earlier studies \cite{PhysRevB.88.085318,PhysRevB.88.075409,PhysRevB.86.115105,PhysRevB.84.045409} only few studies show the out-of-plane dispersion and some show a much larger dispersion for the split valence band along the KH direction with degenerate bands at the H~point \cite{PhysRevB.85.205302,PhysRevLett.105.136805}. 
In Figures \ref{fig:splitting_spinpol}~d)-e) we show that this is an artifact caused by the omission of SOC effects in the calculations.
The out-of-plane dispersion at the $\overline{\text{K}}$~point drops drastically when SOC is included [Fig.~\ref{fig:splitting_spinpol}~d)].
This means that the strong SOC in MoS$_2$ not only gives rise to a splitting of the valence band along the entire KH direction, but it also increases the confinement of these states within the monolayer plane. 
The resulting bands are quasi two-dimensional and therefore very similar to those reported for freestanding monolayers of MoS$_2$ and similar TMDCs \cite{doi:10.1038/nnano.2013.277,PhysRevLett.111.106801}.

In the ARPES spectra in Fig. \ref{fig:all_sbs}~a) the splitting of the valence band at the $\overline{\text{K}}$~point seems obvious.
Since the ARPES spectrum only cuts a 2D plane out of the 3D Brillouin Zone (BZ) the observation of the two sharp bands alone does not prove that they remain split along the entire out-of-plane KH~direction.
However, in photoemission the out-of-plane momentum of the photoelectron can be scanned by varying the photon energy.
Figures \ref{fig:splitting_spinpol}~a)-c) show the ARPES spectra along the high symmetry direction $\overline{\text{KM}}$ for three different photon energies. 
All three spectra represent a part of the 3D-BZ with a different out-of-plane photoelectron momentum k$_\perp$. 
The values for k$_\perp$ were estimated using the free electron final state model and assuming an inner potential of 10 eV.
There is almost no change in the size of the splitting measured with different photon energies, which is a strong indication for a low out-of-plane dispersion and therefore an in-plane confinement as we find it in the DFT calculations. 
To our knowledge there are no other reports regarding an experimental observation of the out-of-plane dispersion of these states in bulk MoS$_2$. 
However, our observation fits very well to recent reports that show the same behavior for WSe$_2$ \cite{10.1038/nphys3105} and WS$_2$ \cite{PhysRevB.91.235202}, which have a very similar band structure.

Figures \ref{fig:splitting_spinpol}~f)~and~g) show the spin-ARPES spectra taken at the $\overline{\text{K}}$~point and the $\overline{\text{K}}$'~point on the opposite side of the BZ. 
One can immediately see that these states are highly spin-polarized and that the polarization reverses at the $\overline{\text{K}}$'~point. 
For the polarization we find $\approx$~65\%  at $\overline{\text{K}}$ and $\approx$~50\%  at $\overline{\text{K}}$'. 
The difference in the size of the polarization is very likely caused by a sightly different position of the beam spot on the sample. 
To reach the $\overline{\text{K}}$'~point, the sample had to be rotated in between the measurements and this could have caused a larger contribution of adjacent terraces. 

An additional mechanism that effectively reduces the observed spin-polarization is photoemission from more than one monolayer, since the polarization reverses from one layer to the next.
Although the low kinetic energy of the photoelectrons of $\approx$~15~-~40~eV makes our method extremely surface sensitive we can estimate the contribution of the individual layers.
Using the universal curve \cite{SIA:SIA740010103}, we assume 5 \AA\ as inelastic mean free path. 
Considering the thickness of one monolayer of MoS$_2$ of $\approx$~6~\AA\ and the photoelectron emission angle of $36^\circ$ for the spin-ARPES measurements, one can estimate that $\approx$~83\% of the photoemission signal comes from the first and the third monolayer and only $\approx$~17\%  from the second. 
In the case of a fully polarized state we would effectively observe only 66\% spin-polarization. 
This simple model suggests that the 65\% spin-polarization that we observed experimentally could in fact result from a polarization close to 100\% in the individual layers. 
Our approximation for the inelastic mean free path is very rough, but a more reliable calculation is not trivial, especially for a highly anisotropic material like MoS$_2$.

In Fig. \ref{fig:splitting_spinpol}~h) we show our results of a 10 monolayer slab calculation including the spin-polarization of the first monolayer. 
For the VBM at the $\overline{\text{K}}$~point we calculated a spin-polarization larger than 99\%. 
This result supports our approximation of the probing depth and the conclusion that the polarization of the individual layers is almost 100\%.
We want to point out that the polarization is not caused by the symmetry breaking at the surface. 
Also in our bulk calculations shown in Fig. \ref{fig:splitting_spinpol}~d) we calculated a spin-polarization larger than 60\% in an individual layer.

So far there are only few publications to which we can compare our results for the spin-polarization. 
The only study of bulk MoS$_2$ so far was published recently by Suzuki and Sakano et al. \cite{Suzuki_Sakano}. 
In this study they present a comparison of centrosymmetric 2H-MoS$_2$ and its non-centrosymmetric polytype 3R-MoS$_2$, which is rare in nature but Suzuki and Sakano et al. were able to grow bulk crystals of both phases. 
Their results seemingly contradict our findings since their spin-ARPES data shows no spin-polarization for 2H-MoS$_2$ and close to 100\%\ for the 3R-type.
They do not directly show the microscopic topography of the 2H-MoS$_2$ samples but an optical micrograph image suggests a much smaller domain size than we find for the exfoliated mineral crystals. 
Therefore it is very likely that the vanishing spin-polarization that Suzuki and Sakano et al. found is a result of the compensation by averaging over a large number of terraces or crystal grains with different orientations.

Our findings are in better agreement with the results for WSe$_2$ that were published by Riley et al. \cite{10.1038/nphys3105}. 
They find close to 100\%\ spin-polarization for centrosymmetric WSe$_2$, which has the same crystal structure and therefore already proves the existence of the hidden spin-polarization in this type of material. 
In the context of this article the essential difference between MoS$_2$ and WSe$_2$ is the larger SOC mediated by tungsten, which is significantly heavier than molybdenum. 
The strength of the SOC directly determines the size of the valence band splitting at the $\overline{\text{K}}$~point. 
We find a splitting of $\approx$~180~meV for MoS$_2$ compared with $\approx$~500~meV that were reported by Riley et al. for WSe$_2$ \cite{10.1038/nphys3105}.
Another result of the larger spin-orbit splitting in WSe$_2$ the topmost valence band at the $\overline{\text{K}}$~point is almost degenerate with the global VBM at the $\overline{\Gamma}$~point. 
However, a larger SOC does not necessarily result in higher spin-polarization: 
E.g. in topological insulators is was found that SOC-induced spin-mixing can effectively reduce the spin-polarization \cite{PhysRevB.86.235106, PhysRevLett.105.266806}. In TMCD's the measured polarization is influenced by the localization of the states in single layers, which is larger in the W-compounds \cite{nature.phys.10.387, PhysRevLett.114.087402}. 
This can explain the higher spin-polarization observed in WSe$_2$ \cite{10.1038/nphys3105}. 
Riley et al. show a dependence of the measured polarization on the photon energy.
They perform a model calculation to consider the contribution of the second monolayer to the photoemission spectrum, that is in good agreement with the observed photon energy dependence. 
We were only able to measure the polarization at one photon energy, which prevents us from performing such analysis but it confirms our assumption that the real spin-polarization in MoS$_2$ is at least $\approx$~65\% but possibly higher.

\subsection*{Conclusions}

Using a combination of ARPES and DFT calculations we were able to show that the distinct valleys in the electronic band structure of bulk MoS$_2$ are quasi two-dimensional and highly spin-polarized despite its three-dimensional, centrosymmetric space group. 
We have experimentally shown the confinement of these electronic states within the plane of the monolayers with the absence of out-of-plane dispersion in the photoemission spectra and confirmed these findings with our DFT calculations.
The spin-polarization that we observed directly in the photoemission experiment was only 65\%. 
However, by taking into account that the probing depth of our method is slightly higher than the thickness of one monolayer, we could estimate that in the individual layers the valleys at the K points are close to being fully polarized. 
This result is also in very good agreement with our DFT calculations, which show a spin-polarization larger than 99\%.

Our findings represent the first observation of this recently introduced type of hidden spin-polarization in MoS$_2$.
This class of materials could broaden the playground for novel spintronic applications as long as the compensated nature of the spin-polarization, both in real and reciprocal space, is taken into account.

\section*{Methods}

\subsection*{Experimental procedures}
\label{exp}

The sample was prepared by cleaving a MoS$_2$ single crystal (SPI Supplies, USA) in air using the scotch tape method creating a surface with shiny and flat areas.
After the cleaving, the samples were immediately put into ultra high vacuum ($< 3 \cdot 10 ^{-10}$ mbar).

The ARPES spectra were measured at two different experimental setups. 
Some spectra were taken using a He discharge source and its spectral lines He I with a photon energy of h$\nu$~=~21.22~eV and He II with h$\nu$~=~40.81~eV.
For this setup we used an MBS A-1 hemispherical analyzer and the sample temperature was kept at $\approx$~15~K. The overall energy resolution was 20~meV.

Further ARPES spectra, including the spin-ARPES spectra, were taken at beamline 5 of DELTA at TU Dortmund.
At DELTA we used a VG Scienta SES-2002 hemispherical analyzer and a Focus SPLEED detector for spin-ARPES measurements \cite{Plucinski2010215}.
We subtracted a Shirley background from the spin resolved spectra and the asymmetry has been corrected for the detector efficiency using a Sherman function of 0.25 and for the non-zero emission angle.
These experiments were performed at room temperature with an overall energy resolution of 100 meV.

The reflectivity measurements were performed using an optical microscope with a tungsten krypton lamp light coupled to optic fibre as an illumination source and with an objective lens with a magnification 10, corresponding to the light spot size of 1~mm diameter.  
The reflected light spectrum was coupled with an optic fibre to a spectrometer with a spectral resolution of 0.5~nm.

\subsection*{Band structure calculations}

All calculations were carried out within the all-electron  full-potential linearized augmented-plane-wave (FLAPW) formalism as implemented in the DFT code {\sc fleur}\cite{fleur}. 
The electron density was determined self-consistently employing the Perdew, Burke and Ernzerhof (PBE) parametrization of the GGA exchange-correlation functional \cite{perdew1996}.
The core electrons were treated fully relativistically by solving the Dirac equation with the spherically averaged effective potential around each nucleus. 
For the valence electrons, space is partitioned into muffin-tin (MT) spheres and an interstitial region. 
In the former we used an angular momentum cutoff $l_{\mathrm{max}}=10$ and 8 for the Mo and S atoms, respectively. 
In the latter, a plane-wave cutoff of 4.2~bohr$^{-1}$ was used for the bulk calculations. 
In the MT spheres, relativistic effects are included in the scalar-relativistic approximation,\cite{koelling1977} while the spin-orbit coupling (SOC) is incorporated self-consistently employing the ``second variation'' technique.
\cite{li1990} An $8 \times 8 \times 4 $~k-point grid was used to sample the bulk BZ. For the film calculations somewhat smaller cutoff (3.8~bohr$^{-1}$) and a $7 \times 7$~k-point grid were used. We used the experimental lattice structures from a publication by Dickinson et al. \cite{dickinson1923}.

\bibliography{LitVerzXSW}

\section*{Acknowledgements}

We want to kindly acknowledge DELTA, Dortmund for providing synchrotron light and the support of Fabian Fritz with the AFM measurements. 
This work was supported by the Helmholtz Association through the Virtual Institute for Topological Insulators (VITI).
G.B. is grateful for computing time on the JUROPA supercomputer at the Jülich Supercomputing Centre (JSC). 

\section*{Author contributions statement}
M.G., L.P., B.K. and C.S. conceived and designed the experiments.
M.G., M.E., E.M., S.D., P.G. and S.C. performed the ARPES measurements.
B.K. performed and analyzed the optical reflectivity measurements. 
M.G. and E.M. performed and analyzed the AFM measurements.
M.G. and L.P. analyzed the photoemission data.
G.B., I.A. and S.B. performed band structure calculations.
M.G., L.P., I.A., G.B. and E.M. contributed to writing of the manuscript.

\begin{figure*}
\includegraphics[width=0.9\linewidth]{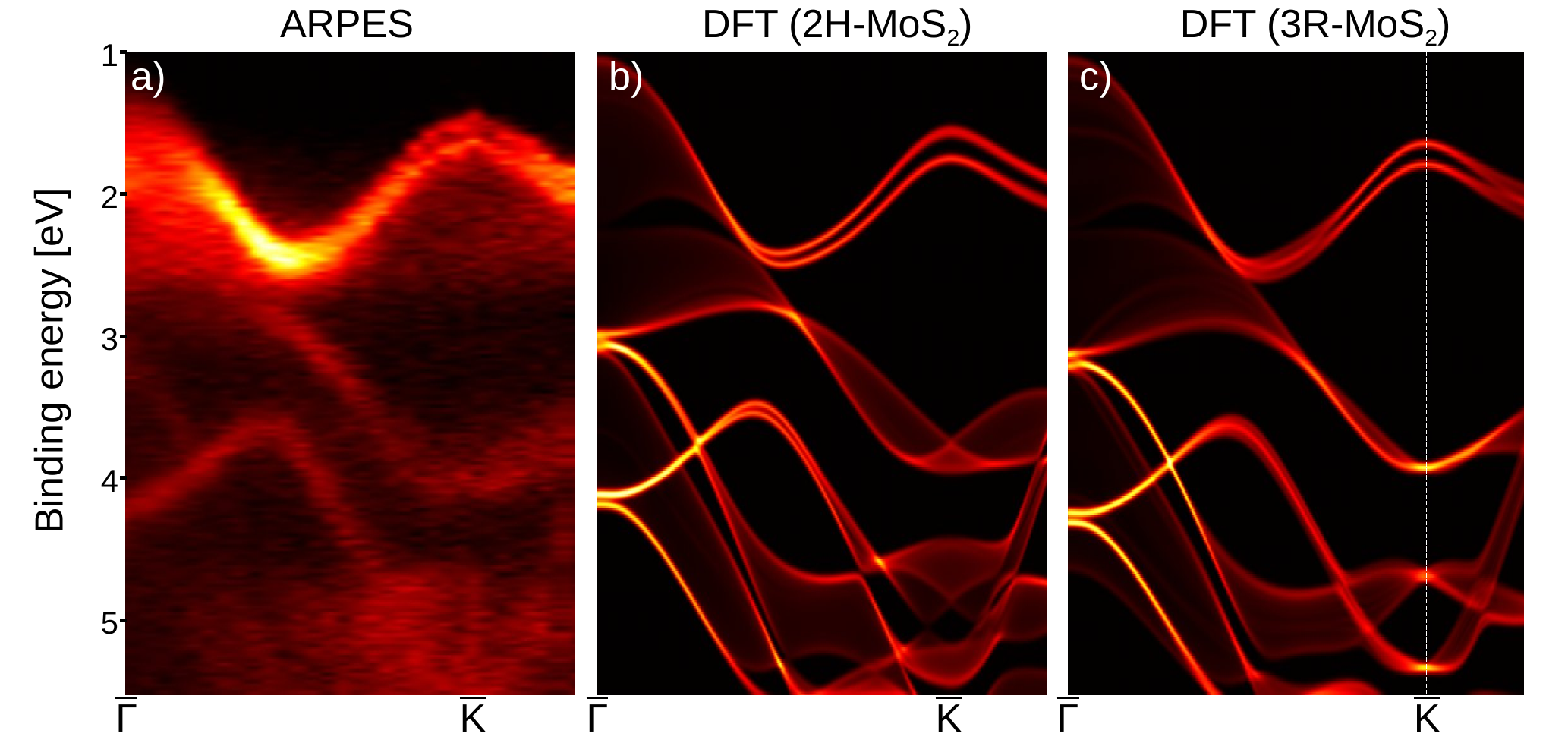}
\caption{\label{fig:ARPES_suppl} 
a) ARPES spectrum along $\overline{\Gamma\text{K}}$~direction measured with h$\nu$~=~47~eV at BL5, DELTA and 
 calculation of bulk projected band structure for 2H-MoS$_2$ in b) and 3R-MoS$_2$ in c).
 In the calculated maps bright areas indicate sharp bands with little out-of-plane dispersion, darker bands a broader projection.
 In b) and c) the Fermi energy is shifted for comparison with a). 
}
\end{figure*}

\newpage

\section*{Supplementary Information}

Figure \ref{fig:ARPES_suppl} shows the comparison of ARPES spectrum along $\overline{\Gamma\text{K}}$~direction to the corresponding calculation of bulk projected band structure for 2H-MoS$_2$ in and 3R-MoS$_2$ in. 
The broadening of the valence band around the $\overline{\Gamma}$~point is a good indication for the delocalization of these states in the out-of-plane~($\Gamma\text{A}$)~direction. 
The high dispersion is a result of the 3D bulk-like character of these states in contrast to the quasi 2D valence band around the $\overline{\text{K}}$~point, which is discussed in the main article.

\end{document}